\documentstyle[12pt]{article}
\begin{document}
\title{
\begin{flushright}
{\small SMI-6-93 }
\end{flushright}
\vspace{0.5cm}
Representations of the compact quantum group $SU_q(N)$
\\and\\geometrical quantization}
\author{
G.E.Arutyunov \thanks{E-mail:$~~$ arut@qft.mian.su}
\thanks{This work is the English translation of the article submitted for
publication in Algebra Analiz}
\\ Steklov Mathematical Institute, Russian Academy of Sciences,\\
Vavilov st.42, GSP-1,117966, Moscow, Russia }
\maketitle
\begin{abstract}
The method of geometrical quantization of symplectic manifolds is applied
to constructing infinite dimensional irreducible unitary representations
of the algebra of functions on the compact quantum group $SU_q(2)$. A
formulation of the method for the general case $SU_q(n)$ is suggested.
\end{abstract} \section
{Introduction} The discovery of new symmetry structures called quantum
groups \cite{Fad}-\cite{Wor} has leaded to the development of the
representation theory of function algebras on quantum groups. In this
article we study the connection of the geometrical quantization with the
representation theory of the function algebra on the compact quantum group
$SU_q(2)$. By geometrical quantization we mean the construction of the
Hilbert space $H$ of quantum states and the observable algebra starting
with the geometry $(M,\omega)$ of a symplectic manifold giving a model of
classical mechanics \cite{K}-\cite{GS}.

The relation of the geometry of the Poisson Lie group $SU(n)$
\cite{S1}-\cite{Sem} with the theory of infinite dimensional
irreducible unitary representations of the function algebra $Fun(SU_q(n))$
was systematically studied by the authors of \cite{WS}-\cite{LS}.
It turned out that these representations are the highest weight representations
and can be parametrized by the elements of Weyl group and points
of maximal torus of $SU(n)$. The aim of this work is to show how infinite
dimensional irreducible unitary representations of $Fun(SU_q(n))$
arise as the result of geometrical quantization in the sense of Kostant
applied to the symlectic leaves of some Poisson structure defined on the
$SU(n)$
group by a classical $r$-matrix. As a simple example, we consider
the case of $SU_q(2)$ and obtain the well-known representations of
 the algebra $Fun(SU_q(n))$ having in mind that the found constructions
keep their value for an arbitrary $n$. Namely, it is due to the fact that
the structure of symplectic leaves in the Poisson Lie group $SU(n)$ is
known \cite{WS1,LS}.

\section{Geometrical quantization}
For constructing representations of the compact quantum groups we need an
analogue of so-called "exact"  quantization in which the Dirac axiom
takes place.

Let us denote ${\cal P}(M,\omega)$ the Poisson algebra of functions
on a symplectic manyfold $(M,\omega)$.
A linear mapping $F\rightarrow \hat{F}$ of the algebra $\cal P$$(M,\omega)$
to the set $B(H)$ of operators, acting in a Hilbert space
$H$ is called Dirac quantization, if it has the following properties
\begin{enumerate}
\item  $\hat{1}=1$,
i.e. the unity on $M$ correspond the unity operator in $H$.
\item $\widehat{\{ F,G\} }=\frac{2\pi i}{h}{[\hat{F},\hat{G}]}~~~$
(Dirac axiom). This means that the mapping $F\rightarrow
\hat{F}$ is a homomorphism of Lie algebras $\cal P$$(M,\omega)\rightarrow
B(H)$.
\item $\widehat{F^*}=\hat{F}^{*}$, i.e. real functions from
$\cal P$$(M,\omega)$ correspond to the self-adjoint operators in $B(H)$.
\item  For some full set $\hat{\cal  P}$ of functions
$F_{1},F_{2},\ldots$ from $\cal P$$(M,\omega)$ the corresponding operators
$\widehat{F_{1}},\widehat{F_{2}},\ldots$ form an irreducible representation
$\hat{\cal  P}$. The set $\hat{\cal  P}$ is called a full, if
any function on $M$, commuting with all $F_i\in \hat{\cal  P}$
with respect to the Poisson brackets is a constant.  \end{enumerate}

A linear mapping $F\rightarrow \check{F}$, obeying only
(1)-(3) is said to be prequantization. For a given symplectic manifold
$(M,\omega)$ prequantization can be constructed as follows.
\cite{K}. Let us cover the manifold $M$ by open covers $U_{\alpha}$ in
such a way as to obtain in every cover $U_{\alpha}$ the identity
$\omega=d\theta _{\alpha}$ for a suitable 1-form $\theta _{\alpha}$ in
$U_{\alpha}$.  Define in $U_{\alpha}$ the operator:  \begin{displaymath}
\check{F}_{\alpha}=F_{\alpha}+\frac{h}{2\pi
i}\xi_{F}-\theta_{\alpha}(\xi_{F})~,
\end{displaymath}
that acts on the space $C^{\infty }(U_{\alpha})$.
Here $F_{\alpha}\in C^{\infty }(U_{\alpha})$ and
$\xi_{F}$-is a hamiltonian vector field with the generating function $F$,
considering as an operator in $C^{\infty}(M)$.

If the form $\omega$ defines an integer cohomology class in $H^2(X,R)$,
then according to fundamental Kostant's theorem, the introduced
local operators $\check{F}_{\alpha}$ can be composed in a one global
operator $\check{F}$, acting on the space of smooth sections of a linear
bundle $L$ over $M$. The form $\theta_{\alpha}$ can be realized as a local
expression for some connection $\bigtriangledown$ of the bundle $L$ in the
trivialization $U_{\alpha}$. Let us note that in the case of real $\omega$
there exists on $L$ a hermitian structure compatible with the connection
$\nabla $, and the prequantization operator $\check{F}$ has the following
form:  \begin{displaymath} \check{F}=F+\frac{h}{2\pi
i}\bigtriangledown_{\xi_{F}}~~, \end{displaymath} GDE $\nabla
_{\xi_{F}}=\xi_{F}-\frac{2\pi i}{h}\theta(\xi_{F})$.

The prequantization space, i.e. the space of sections $\Gamma(L,M)$
of $L$ over $M$, is too large to fulfill the condition (4).
The subspace $\Gamma(L,M,P)$, on which an irreducible representation
of the observable algebra acts are distinguished from $\Gamma(L,M)$ by
means
of a real or complex polarization $P$
\cite{Kir},\cite{GS}. Namely, the elements of $\Gamma(L,M,P)$ are
the smooth functions $s\in\Gamma(L,M)$, obeying the equation
$$\nabla_{\eta}s=0$$ for any vector field $\eta\in P$.

It should be noted that for the case in question prequantization can
be thought of as the assignment of homomorphism of the Poisson algebra
$\cal P$$(M,\omega)$ in some abstract associative algebra $\cal A$ with
involution (any associative algebra with respect to product can be
turned in to the Lie algebra by introducing an ordinary commutator).
The quantization procedure is realized then as the construction of
irreducible representations of the algebra $\cal A$.

Let us consider now the algebra of regular functions on the compact
quantum group $SU_q(2)$.  This is the associative algebra
$Fun(SU_q(2))$ with unity and involution $*$, generated by
two  generators $t_{11}$, $t_{21}$ modulo the relations:
\begin{displaymath}
t_{11}t_{21}^{*}=qt_{21}^{*}t_{11}~~,~~t_{21}t_{21}^{*}=t_{21}^{*}t_{21}~~,
\end{displaymath}
\begin{equation}
t_{11}t_{11}^{*}-t_{11}^{*}t_{11}=(1-q^2)t_{21}t_{21}^{*}~~,~~ \label {1}
\end{equation}
\begin{displaymath}
t_{11}^{*}t_{11}+t_{21}t_{21}^{*}=1~~,~~t_{11}t_{21}=qt_{21}t_{11}~.
\end{displaymath}
where $q$ is the deformation parameter. The algebra $Fun(SU_q(2))$ has
the Hopf algebra structure. It becomes commutative in the limit $q=1$
and can be identified with the algebra of commutative polynomials
generated by matrix elements of the fundamental representation
of $SU(2)$, i.e. as an algebra of regular functions on
$SU(2)$. Introduce a new parameter $h$ (Plank's constant) connecting
with $q$ as $q=e^{-h}$. Let $h$ belongs to the interval $E=(0,a]$ where
$a$ is some positive constant. Then the relations
(\ref{1}) give the family $A_{h}$ of associative algebras supplied with
involution:
\begin{displaymath}
t_{11}t_{21}^{*}=e^{-h}t_{21}^{*}t_{11}~~,~~t_{21}t_{21}^{*}=t_{21}^{*}t_{21}~~,~~
\end{displaymath}
\begin{equation}
t_{11}t_{11}^{*}-t_{11}^{*}t_{11}=(1-e^{-2h})t_{21}t_{21}^{*}~~,~~  \label{2}
\end{equation}
\begin{displaymath}
t_{11}t_{21}=e^{-h}t_{21}t_{11}~~,~~
t_{11}^{*}t_{11}+t_{21}t_{21}^{*}=1~~.~~
\end{displaymath}

It is well known that $SU(2)$ is a Poisson Lie group.
\cite{D}. In other words, the function algebra
$A_{0}=Fun(SU(2))$ is a Poisson Hopf algebra (i.e. $A_{0}$
has the Hopf as well as the Poisson algebra structure with
one and the same product with coproduct
$A_{0}\rightarrow A_{0}\otimes A_{0}$ being homomorphism of
Poisson algebras).

The family (\ref {2}) of the algebras $A_{h}$ give
a deformation, or in other words, quantization of the Poisson algebra
$A_{0}~$ in the sense of associative Hopf algebras \cite{D}, i.e. for any
$h\in E$ $A_h$ has the Hopf algebra which goes to the structure of $A_{0}$
when $h\rightarrow 0$. In this case the correspondence principle claims
that \begin{equation} \{ F,G\} =\lim _{h\rightarrow
 0}\frac{1}{h}[\hat{F},~\hat{G}]~~,        \label{3} \end{equation} where
we have the Poisson brackets of any functions $F$ and $G$ from $A_{0}$ on the
left and the commutator of their images in $A_{h}$ on the right.
The Poisson brackets on $A_{0}$ arising in such a way are quadratic ones.
Thus the construction of the family $A_{h}$ can be regarded
as quantization of the quadratic brackets on the Poisson Lie group
$SU(2)$.

Let us note that the defining relations (\ref{2}) of the noncommutative
algebra $A_{h}$ for the fixed $h$ are specified by
a quantum $R$-matrix being a matrix solution of the quantum Yang-Baxter
equation (QYBE) and quadratic brackets on $SU(2)$
compatible with the Hopf algebra structure are constructed by using
the canonical $r$-matrix:
\begin{displaymath}
r=\frac{1}{2}\sum_{\alpha>0}X_{\alpha}\otimes
X_{-\alpha}-X_{-\alpha}\otimes X_{\alpha}
\end{displaymath}
giving a solution of the classical Yang-Baxter equation
(CYBE) \cite{D,S1}. Here $X_{\alpha}$ is
a basis of positive roots of $su(2)$. Supposing
$R$-matrix to be decomposed in the $h$-series:  \begin{displaymath}
R=1+hr+O(h^2) \end{displaymath} in the second order of the QYBE
we obtain for $r$ the CYBE. It is the connection between the QYBE and the CYBE
that governs by the correspondence principle (\ref{3}).

Let us consider now  $*$-representations $\pi$ of the algebra (\ref{2}) when
$h\in
E$ in  separable Hilbert space. As it is known \cite{WS},
every irreducible *-representation $\pi$ of the algebra (\ref{2}) in
Hilbert space $H$ is unitary equivalent to the one of the following two
series:  \begin{enumerate} \item
One dimensional representations  $\xi_{\psi}$ given by the formulae
\begin{displaymath}
      \xi_{\psi}(t_{11})=e^{i\psi}~~~~~,~~~~~\xi_{\psi}(t_{21})=0~~,~~\psi\in
  R/{2\pi Z}~~.
\end{displaymath}
\item Infi\-ni\-te dimensional representations $~~\rho_{\psi}$ having in
an ortho\-normal bases $\{ e_k\} _{k=0}^{\infty}$ the form:
   \begin{equation}
   \rho_{\psi}(t_{11})e_0=0~~,~~\rho_{\psi}(t_{11})e_k=\sqrt{1-e^{-2kh}}e_{k-1}
  ~~,~~\rho_{\psi}(t_{21})e_k=e^{i\psi-kh}e_k       \label{4}
\end{equation}
\end{enumerate}

Let us describe now the procedure of finding the connection between infinite
dimensional representations (\ref{4}) of the algebra (\ref{2}) and the
geometry of $SU(2)$. In the limit $h\rightarrow 0$ the defining relations
of the algebra (\ref{2}) give a degenerate Poisson structure on $SU(2)$.
The reduction of this structure on it's symplectic leave of a maximal
dimension generates a symplectic form $\omega$. We will regard now the
symplectic leave with the form $\omega$ as a model of classical
mechanics. To construct a Hilbert space $H$ and a set of operators on it
giving a quantum analogue of the system in question one needs to show the
absence of cohomology obstacles. Namely, the cohomology class of the form
$\omega$ must be integer\cite{K}.  Further we will construct Hilbert space
$H$ and the representation of the maximal commutative subalgebra in
the Poisson algebra $Fun(SU(2))$ and point out it's relation
with the representation (\ref{4}) of the algebra (\ref{2}).
\section{Geometrical quantization of symplectic leaves in $SU(2)$}
\setcounter{equation}{0} Consider the limit $h\rightarrow 0$. The
defining relations of the algebra (\ref{2}) produce the following
quadratic Poisson brackets on $Fun(SU(2))$:

\begin{displaymath} \{ t_{11},~t_{21}\} =t_{11}t_{21}~~,~~
\{ t_{11},~\bar{t_{11}}\} =t_{21}\bar{t_{21}}~~,~~ \end{displaymath}
\begin{equation} \{ t_{11},~\bar{t_{21}}\} =-t_{11}\bar{t_{21}}~~,~~ \{
t_{11},~\bar{t_{21}}\} =0~~,~~                           \label{5}
\end{equation} \begin{displaymath} \{ t_{21},~\bar{t_{11}}\}
=-t_{21}\bar{t_{11}}~~,~~ \{ t_{11},~\bar{t_{11}}\} =-t_{21}\bar{t_{11}}~~,~~
\end{displaymath}
where the element $g\in SU(2)$ is parametrized by $t_{11},t_{21}$:
\begin{displaymath}
g=\left(\begin{array}{rr}t_{11}&-\bar{t_{21}}\\t_{21}&\bar{t_{11}}\end{array}\right)
\end{displaymath}
obeying the condition $\mid t_{11}\mid ^2+\mid t_{21}\mid ^2=1$, and the
bar means the complex conjugation. Brackets (\ref{5}) generate corresponding
to the coordinate functions strictly hamiltonian vector fields on
$SU(2)$:  \begin{displaymath}
\xi_{t_{11}}=-t_{11}\left(t_{21}\frac{\partial}{\partial
t_{21}}+\bar{t_{21}}\frac{\partial }{\partial
\bar{t_{21}}}\right)+2t_{21}\bar{t_{21}}\frac{\partial }{\partial
\bar{t_{11}}}~~,
\end{displaymath}
\begin{displaymath}
\xi _{t_{21}}=t_{21}\left(-\bar{t_{11}}\frac{\partial
}{\partial \bar{t_{11}}}+t_{11}\frac{\partial }{\partial t_{11}}\right)~~.
\end{displaymath}
Moreover, we have
$\bar{\xi _{t_{11}}}=-\xi_{\bar{t_{11}}}$ and
$\bar{\xi _{t_{21}}}=-\xi_{\bar{t_{21}}}$.

The algebra $Fun(SU(2))$ is supplied with the involution that is the
complex conjugation. The quadratic brackets (\ref{5}) are antiinvolutive
in opposite to the ordinary brackets on $T^*M$. For example,
\begin{displaymath} \bar{  \{ t_{11},t_{21}\} }=-\{
       \bar{t_{11}},\bar{t_{21}}\}~~.  \end{displaymath}
Thus the reduction of the brackets (\ref{5}) on a symplectic leave
gives a symplectic form $\omega $ having the property $\bar{\omega }=-\omega$,
that means the absence of it's real part.  Therefore, for the real
function $F~~(F=\bar{F})$ the strictly hamiltonian vector field
$\xi_{F}$, defined by equation
\begin{displaymath}
2\omega(\xi_{F},\ldots)+dF=0 \end{displaymath} obeys the condition:
$\xi_{F}=-\bar{\xi_{F}}$.

Such behavior of brackets (\ref{5}) with respect to the unvolution gives
a hint that in the geometrical quantization procedure we have to require
this time instead of
\begin{displaymath} \frac{2\pi i}{h}[\check{F},\check{G}]=\check{\{ F,G\}
} \end{displaymath} then fulfillment of the following relation:
\begin{equation} \frac{2\pi }{h}[\check{F},\check{G}]=\check{\{ F,G\} }~~.
\label{8} \end{equation} Then the choice of prequantization operators
in the form:  \begin{equation}
\check{F}=F+\frac{h}{2\pi }\xi_{F}-\theta(\xi_{F}) \label{kvan}
\end{equation} obeys automatically (\ref{8}). Here
$\theta$ is such  1-form that $\omega=d\theta$. Since
$\omega$ is imaginary and for real functions $F$:
$\xi_{F}=-\bar{\xi_{F}}$, the form $\theta$ can be chosen to be imaginary. Then
\begin{displaymath} \bar{\theta(\xi_{F})}=\bar{\theta
}(\bar{\xi_{F}})=-\theta(-\xi_{F})=\theta(\xi_{F}), \end{displaymath} i.e.
the function $\theta(\xi_{F})$ is real for the real $F$. Thus
we have the possibility to realize the prequantization operators in the
space of finite functions on a symplectic leave $M$. By introduction of
the scalar product:  \begin{equation} (\varphi _1,\varphi
_2)=\int _{M}\varphi _1\bar{\varphi _2}\omega~~, \end{equation}
where $\omega$ is the Liouville measure on the leave $M$, this space
is turned to be Hilbert. (For $SU(2)$ the maximal dimension of symplectic
leaves is equal to 2). The prequantization operators corresponding
to real finite functions are symmetrical and essentially self-adjoint.

Let us find explicitly the symplectic form $\omega$ on the symplectic leave
$M$ of the quadratic brackets (\ref{5}). Hamiltonian vector fields
of the coordinate functions annihilate the following two functions:
\begin{displaymath}
        c_1=\mid
 t_{11}\mid ^2+\mid t_{21}\mid ^2
{}~~,~~c_2=(t_{21}/\bar{t_{21}})^{1/2}~~`~~\mid t_{21}\mid \neq0 ~~.
\end{displaymath} Therefore if $\mid t_{21}\mid \neq0$
the symplectic leave consists of the matrices:
\begin{displaymath}
M_{\psi}=\left[g:~~
g=\left(\begin{array}{rr}t_{11}&-\bar{t_{21}}\\t_{21}&\bar{t_{11}}\end{array}\right)
\in SU(2),~~ Arg~t_{21}=\psi,~~ \mid t_{21}\mid \neq0 ~\right]~~,
\end{displaymath}
where $0<\psi\leq 2\pi$. $M_{\psi}$ is described by the one complex coordinate
$z=t_{11}$ obeying the condition $z\leq 1$, since $\mid t_{11}\mid
  ^2+\mid t_{21}\mid ^2=1$ and $\mid t_{21}\mid >0$. So we see that $M_{\psi}$
is an open circle of the radius one on a complex plane, i.e. a noncompact
complex manifold. The range of brackets is constant for any point inside
the round. Any differential 2-form
$\omega$ on $M_{\psi}$ is:  \begin{displaymath}
\omega_{\psi}=a(t_{11},\bar{t_{11}},\psi)dt_{11}\wedge d\bar{t_{11}}~.
\end{displaymath}
The condition $\bar{\omega }=-\omega$ gives that $a$ is real, i.E.
$a(t_{11},\bar{t_{11}},\psi)\in R$. The reduction of the Poisson brackets on
the leave $M_{\psi}$ should coincide with $\omega_{\psi}$:
\begin{displaymath}
\{ F,G\} =2\omega_{\psi}(\xi_{F},\xi_{G})~.
\end{displaymath}
 From here we find:
\begin{equation}
\omega=\frac{1}{2(1-\mid z\mid^{2})}dz\wedge d\bar{z}~~.
\end{equation}

Let $X=M_{\psi}/\{ 0\} $ be
the open circle on the complex plane without zero.
Then there exists on $X$ the 1-form $\theta$ defined by:  \begin{equation}
\theta=\frac{1}{4}\ln(1-\mid z\mid^{2})(\frac{dz}{z}-\frac{d\bar{z}}{\bar{z}})=
\frac{i}{2}\ln(1-\rho ^{2})d\varphi~,
\end{equation}
for which $\omega=d\theta$. Here $z=\rho e^{i\varphi}~,~\varphi \in
S^1~,~0<\rho<1$.

Denote $L(X)\approx X\times C$  a trivial linear bundle over $X$.
On $C$ one can choose the usual hermitian structure. Since $L(X)$  is trivial
it has the unit section $s$ globally determined on $X$. Let us think of
$\theta$ to be a connection form $\tilde{\theta}$
of the bundle $L(X)$, corresponding to the section $s$, i.E.
$$s^{*}\theta =\tilde{\theta}~.$$
The curvature $\Omega$ of this connection
is also defined on the base $X$ by means of $s$. It coincides with
$\frac{1}{h}\omega$. Note, that the connection introduced in such a way
is not flat but the cohomology class corresponding to
$\omega$ in $H^2(X,R)$ is equal to zero. Thus the cohomology obstacles
are absent.

Let us construct the prequantization operator for the coordinate function
$t_{21}$, using the formula (\ref{kvan}):  \begin{equation}
\check{t_{21}}=t_{21}+\frac{h}{2\pi}\xi _{t_{21}}-\theta(\xi _{t_{21}})~~.
\end{equation} In the $(\rho,\varphi)$ coordinates we have:
\begin{equation} \check{t_{21}}=t_{21}+t_{21}\left(\frac{h}{2\pi i}
\frac{\partial }{\partial \varphi}-\frac{1}{2}\ln(1-\rho ^{2})\right)~~.
\end{equation}
Consider the polarization generated by the vector field $\eta=\frac{t_{21}}{i}
\frac{\partial }{\partial \varphi}$.
The polarization is real $(\eta=-\bar{\eta})$ and it's leaves on $X$ are
circles. The quantization space $H$ associated with the polarization
$\eta$ consists of sections $s(\rho,\varphi)$ of the linear bundle $L(X)$
over $X$ that are horizontal ones with respect to the direction defined by
the vectr field $\eta$.
Thus $H$ is modeled by solutions of the equation $\nabla
_{\eta}s=0$. Being rewritten in the variables $(\rho,\varphi)$ it takes
the form:  \begin{equation} \frac{h}{i\pi } \frac{\partial s}{\partial
\varphi}-\ln(1-\rho ^{2})s=0.  \label{we} \end{equation} Trying to solve
this equation we run into the well-known problem
of geometrical quantization
\cite{Ki,GS}. As was mentioned above the leaves of the
polarization are circles, i.e. non simply connected manifolds.
Parallel transport of a section along the closed way $\gamma
$ laying in the leave leads to the multiplication of $\gamma$ on
$$Q(\gamma)=e^{\frac{2\pi}{h}\int_{\gamma}\theta}.$$ Since,
generally speaking, an arbitrary
way $\gamma$ can not be squeezed to a point the number $Q(\gamma)$
is not equal to one. Clearly, in this case any solution of equation
(\ref{we}) equals zero on the leave considered.
Denote $\hat{X}$ the set of polarization leaves for which
$Q(\gamma)=1$.  The set $\hat{X}$ is known as Bohr-Zommerfeld
submanifold \cite{Kir}.
The way out of the problem consists in considering instead of
$\Gamma (L,X,P)$ the set of distributions being solutions of equations
(\ref{we}).  All solutions of such a type (up to a multiplication constant
) have the form:  \begin{equation}
s_{n}(\rho,\varphi)=e^{in\varphi}\delta
\left(\rho-\sqrt{1-e^{\frac{nh}{\pi}}}\right) \end{equation} and their support
coincides with $~~\hat{X}$ (the uni\-ty of circles).  In the variables
parametrizing points of the symplectic leave $M_{\phi}$, where $\phi\in
S^{1}$, the coordinate function $t_{21}$ is written as follows:
\begin{equation} t_{21}=\mid t_{21}\mid e^{i\phi}=\sqrt{1-\rho
^{2}}e^{i\phi}~. \end{equation} Let us introduce the set of elements
$s_{n}^{\psi}(\rho,\varphi,\phi)$ numerating by the parameter $\psi\in S^1$:
\begin{equation} s_{n}^{\psi}(\rho,\varphi,\phi)= e^{in\varphi}\delta
\left(\rho-\sqrt{1-e^{\frac{nh}{\pi}}}\right) \otimes \delta
\left(\phi-\psi\right)~~.  \end{equation} Then the operator $\check{t_{21}}$
acts on the elements of $s_{n}^{\psi}$ as the multiplication on the
function $t_{21}$:  \begin{displaymath} \check{t_{21}}=\sqrt{1-\rho
 ^{2}}e^{i\phi}~.  \end{displaymath} Value
s of the parameter $\psi $
are in correspondence with points of a maximal torus of the group $SU(2)$.
Introduction of the space $\{ s_{n}^{\psi}(\rho,\varphi,\phi)\} $
means the choice of some special quantization procedure of the function
algebra on $SU(2)$ under which the functions depending only on the torus
coordinates realized by multiplication operators.
We find
\begin{equation}
\check{t_{21}}s_{n}^{\psi}(\rho,\varphi,\phi)=e^{i\psi+\frac{nh}{2\pi}}
s_{n}^{\psi}(\rho,\varphi,\phi).                 \label{15}
\end{equation}
The natural requirement $1-e^{\frac{nh}{\pi}}\geq 0$ defines an allowed
region of $n$ to take it's value:  $n\in N=\{
0,-1,-2,\ldots \} $. Thus in the representation $s_{n}^{\psi}$
the operator $\check{t_{21}}$ has the spectrum:  $e^{i\psi+\frac{nh}{2\pi}}$,
$n\in N$ that is identical to the one of the operator
$\rho_{\psi}(t_{21})$ in the representation (\ref{4}) when
$q=e^{-\frac{h}{2\pi}}$.

By analogy we obtain for the operator $\check{\bar{t_{21}}}$:
\begin{equation}
\check{\bar{t_{21}}}s_{n}^{\psi}(\rho,\varphi,\phi)=e^{-i\psi+\frac{nh}{2\pi}}
s_{n}^{\psi}(\rho,\varphi,\phi)~.                    \label{16}
\end{equation}

The straightforward calculations of the operators corresponding to the
coordinate functions $t_{11}$ and $\bar{t_{11}}$ show that
they do not preserve the quantization space $H_{\psi}$. However, let us
note that the coordinate functions  $t_{21}$, $\bar{t_{21}}$ and 1 generate
the maximal commutative subalgebra with respect to the Poisson brackets.
Then the operators $\check{t_{21}},\check{\bar{t_{21}}}$ and 1 can be
considered
as the representation of this algebra in the space of horizontal sections
$H_{\psi}$ over Bohr-Zommerfeld submanifold.
We will show that this representation can be prolonged to representations
of the algebra (\ref{2}) in $H_{\psi}$ that unitary equivalent to (\ref{4}).

Introduce in the space  $H_{\psi}$
a scalar product $(,)$ in such a way as to obtain $(s_{n},s_{m})=\delta _{nm}$.
Let the operator $\check{t_{11}}$ has the matrix
$\cal A$=$\parallel a\parallel_{nm}$ in the bases $s_{n}^{\psi}$, i.E.:
\begin{displaymath} \check{t_{11}}s_{n}^{\psi}=\sum
_{m}a_{nm}s_{m}^{\psi}~~. \end{displaymath} Then \begin{displaymath}
\check{\bar{t_{11}}}s_{n}^{\psi}=\sum _{m}a_{mn}^{*}s_{m}^{\psi}~~.
\end{displaymath} The relations (\ref{2}) combined with (\ref{15}) and
(\ref{16}) give us:  \begin{equation}
\check{\bar{t_{11}}}\check{t_{11}}s_{n}^{\psi}=
(1-e^{\frac{nh}{\pi}})s_{n}^{\psi}                         \label{17}
\end{equation}
and
\begin{equation}
\sum _{m}e^{\frac{nh}{2\pi}}a_{nm}s_{m}^{\psi}=
\sum _{m}e^{(m-1)\frac{nh}{2\pi}}a_{nm}s_{m}^{\psi}~~.         \label{18}
\end{equation}
 From the last equation one has:
\begin{displaymath}
\check{t_{11}}s_{n}^{\psi}=b_{n}s_{n+1}^{\psi}~,
\end{displaymath}
\begin{equation}
\check{\bar{t_{11}}}s_{n}^{\psi}=b_{n-1}^{*}s_{n-1}^{\psi}~,
\label{19} \end{equation} where $b _n$ are coefficients that should be defined.
Substituting (\ref{19}) in (\ref{17}) we find: $$b_n=
e^{i\phi _{n}}\sqrt{1-e^{\frac{nh}{\pi}}}~~,~~\phi _n\in S^1~~,~~n<0~.$$

Let us show now that the representation $\hat{T}$
characterized by the infinite dimensional set $\{ \phi _1,\phi _2,\ldots\} )$
PREDSTAWLENIE          :  \begin{displaymath}
\check{t_{21}}s_{n}^{\psi}(\rho,\varphi,\phi)=e^{i\psi-\frac{nh}{2\pi}}
s_{n}^{\psi}(\rho,\varphi,\phi)~,
\end{displaymath}
\begin{displaymath}
\check{\bar{t_{21}}}s_{n}^{\psi}(\rho,\varphi,\phi)=e^{-i\psi-\frac{nh}{2\pi}}
s_{n}^{\psi}(\rho,\varphi,\phi)~,
\end{displaymath}
\begin{displaymath}
\check{t_{11}}s_{n}^{\psi}(\rho,\varphi,\phi)=e^{i\phi_n}\sqrt{1-e^{-\frac{nh}{\pi}}}s_{n-1}^{\psi}(\rho,\varphi,\phi)~, \end{displaymath}
\begin{displaymath}
\check{\bar{t_{11}}}s_{n}^{\psi}(\rho,\varphi,\phi)=
e^{i\phi
_{n+1}}\sqrt{1-e^{-\frac{(n+1)h}{\pi}}}s_{n+1}^{\psi}(\rho,\varphi,\phi)~,~n\geq 0
\end{displaymath}
is unitary equivalent to the representation (\ref{4}) denoted by $T$.
Consider the operator $U$ acting on $s_{n}^{\psi}$ in the following manner:
$$Us_{n}^{\psi}=e^{i\sum_{k=1}^{n-1}\phi _k}s_{n}^{\psi}~~,~~Us_{0}=0~~.$$
Clearly  $U$ is unitary and $U~\hat{T}U=T$, i.E. $T_{\phi}$ and $T$ are
equivalent.

\section{Geometrical quantization of the Poisson Lie group $SU_q(n)$}
\setcounter{equation}{0} Here we present the general formulation of the found
on the $SU_q(2)$ example the connection of geometrical quantization with
the theory of unitary infinite dimensional representations an arbitrary
compact quantum group.

Let $Fun(G_{q})$ be the function algebra on a compact quantum group $G_{q}$.
When the deformation parameter $q$ goes to one $G_{q}$ turns
into the Poisson Lie group $G$. The Poisson structure arising in such a way
is degenerate. It's symplectic leaves are parametrized by the elements of
Weyl group and by points of maximal torus
$H$ of the group $G$. Let us reduce the Poisson brackets on the leave  $M$ and
consider in the Poisson algebra the maximal commutative subalgebra $B$
with respect to the Poisson brackets. Geometrical quantization gives the
realization  $\pi $ of this subalgebra by differential operators of the
first order acting in the space  $\Gamma
(L,M)$ of sections of a linear bundle $L$ over $M$.  $\Gamma (L,M)$
can be identified with the space of prequantization.  Let us chose such a
polarization $P$ that all operators $\pi (B)$ are
diagonal in the corresponding space $\Gamma (L,M,P)$. Since the algebra
$B$ is commutative then the polarization with the desired property exists.
The space $\Gamma (L,M,P)$ are defined to be the set of distributions
horizontal in the direction of the polarization.
In $\Gamma (L,M,P)$ the irreducible representation of the function algebra
$Fun(G_{q})$ acts. It's reduction on the image of $B$ in $Fun(G_{q})$
coincides with $\pi (B)$.  Thus geometrical quantization of leaves of
Poisson structure gives an explicit realization of
the space of the corresponding representation of the quantum
group and the realization of $\pi (B)$
the algebra $B$ on this space.  Since irreducible infinite dimensional
unitary representations of $Fun(G_{q})$ are representations with highest
weight they are defined up to the unitary equivalence by restrictions on
the maximal commutative subalgebra
$B$ (an analogue of Cartan subalgebra).
\section{Concluding remarks}
Note that Dirac axiom plays a crusual role in the considered example. The
reason for which the standard geometrical quantization of the Poisson
algebra $\cal P$$(M,\omega)$ does not immediately give representations of
a quantum group as follows. The deformation procedure of Hopf algebras is
not "exact" quantization. In general, for any two observables one has
$\cal P$$(M,\omega)$ the equality:
$$[\hat{F},\hat{G}]=\frac{h}{2\pi}\widehat{\{ F,G\}
}+O(h^{2}).$$ Existence of the hole series in Plank's constant on the left
hand side is the well known problem arising in quantization of nonlinear
(in particular quadratic) Poisson brackets.
Namely, one can not substitute Poisson brackets by an operator analog
because it is unknown how to order their right hand side.
The choice of some ordering leads to the appearance of highest powers of $h$
in the equality under consideration. Thus, in the case of nonlinear
brackets we have only some peculiar subalgebras (in our example it is a
commutative algebra) for which the series on the right vanishes, the
quantization becomes "exact" and can be formulated purely in terms of geometry.
Hence the natural question arises of whether it is possible to extend somehow
the class of subalgebras in $\cal P$$(M,\omega)$
being the subject of quantization. Clearly, existence of such an extension
is connected with the possibility to generalize the usual procedure of
geometrical quantization. This should be done in a way as to realize
a quantum group as the Hopf algebra of invariant differential operators
(but not the first order). The positive solution of this problem can be
applied then to the construction of dynamical systems on quantum groups
\cite{AV},\cite{SW}.
$$~~$$
{\bf ACKNOWLEDGMENT}
$$~$$
The author would like to express his gratitude to I.Ya.Aref'eva,
I.V.Volovich and P.B.Medvedev for interesting discussions.  $$~$$
 \end{document}